\documentclass{article}
\usepackage{PRIMEarxiv}
\usepackage[utf8]{inputenc} 
\usepackage[T1]{fontenc}    
\usepackage{hyperref}       
\usepackage{url}            
\usepackage{booktabs}       
\usepackage{amsfonts}       
\usepackage{nicefrac}       
\usepackage{microtype}      
\usepackage{lipsum}
\usepackage{fancyhdr}       
\usepackage{graphicx}       
\graphicspath{{media/}}     

\usepackage{cite}
\usepackage{verbatim, amsmath, amsfonts, amssymb, amsthm} %
\usepackage{systeme}
\usepackage{url}
 \usepackage{bbm}
\usepackage{nicefrac} 
\theoremstyle{thmstyleone}%
\newtheorem{theorem}{Theorem}
\theoremstyle{thmstyletwo}%
\theoremstyle{thmstylethree}%
\title{Option Pricing And CVA Calculations Using the Monte Carlo-Tree (MC-Tree) Method}
\author{
Yen Thuan TRINH \\
  School of Mathematical Sciences\\
  University College Cork\\
  Ireland \\
  \texttt{trinhthuanyen2017@gmail.com} \\
   \And
 Bernard HANZON, Chair of Mathematics \\
  School of Mathematical Sciences\\
  University College Cork\\
  Ireland \\
  \texttt{b.hanzon@ucc.ie } \\ 
}

\begin{document}
\fontsize{13}{17}\selectfont
\maketitle

\begin{abstract}
The binomial tree method and the Monte Carlo (MC) method  are popular methods for solving option pricing problems. There is need for faster and more accurate option price calculations. We introduce a new method, the  MC-Tree method, that combines the MC method with the well-known recombining binomial tree based on Pascal's triangle for pricing single asset options. Our approach uses a mixing distribution on the tree, for which we obtain the corresponding compound distribution on the tree outcome. As well known in the literature, the standard Gaussian distribution is the distribution with the maximal entropy among distributions with zero mean and unit variance. The compound density that we obtain is not exactly equal the Gaussian density, but has very high entropy. Also we introduce techniques to correct for the deviation from the Gaussian. Based on these techniques we develop an algorithm for calculation of the Credit Valuation Adjustment (CVA) on an American put option, a challenge in computation due to the complexity of the American options and modelling CVA. We also present numerical results. Based on our these, the MC-Tree method is more accurate than the well-known Least Square Monte Carlo method (LSM) for American option valuation  proposed by Longstaff Schwartz (2001) at the same numbers of simulations. The MC-Tree method performs better than other methods: Cox, Ross Rubenstein (CRR) and Jarrow-Rudd(JR) in
terms of accuracy, using the same tree depth. Also, the MC-Tree method with the distribution correction technique dramatically improved the accuracy, resulting in the practically exact solutions, compared to analytical solutions, at the tree depth N=50 or 100 and MC-drawings M=100000. The bias-correction technique makes the resulting tree model complete in the sense of financial mathematics and obtains the risk-neutral probability.

\end{abstract}

\maketitle
\section{Introduction}\label{sec1}
The pioneering work of Bachelier (1900) is now seen as the forerunner of what would become a massive usage of mathematical models in finance since the last quarter of the 20th century. 
There is a demand for 
fast and accurate methods for pricing options when more exotic financial instruments are developed and traded on the market over the years.  Black and Scholes (see \cite{Black&Scholes}) derived the Black-Scholes equation to price derivatives on a single asset. 
Subsequently analytical solutions for quite a number of derivatives have been derived, using either the Black-Scholes partial differential equation or the discounted risk-neutral expectation method, but it is too difficult to solve the equation analytically for an arbitrary derivative. Based on the Black-Scholes model, many methods for option valuation have been developed, but there is still room for improvement. Our contribution is to introduce a new computational approach 
to find the value of financial derivatives with arbitrary boundary conditions and generalization to the case of American options and other classes of options such as Barrier options.\\
 Boyle \cite{Boyle1977} introduced the Monte Carlo approach to option pricing, and it is still very popular because of its flexibility to approximate all kinds of option prices. Monte Carlo simulation has been extended in pricing American options with popular methods such as the Stochastic Mesh \cite{broadie2004stochastic}, the Least Square Monte Carlo \cite{Longstaff}, and the State-Space partition \cite{tilley1995valuing}, and so on. 
Longstaff and Schwartz in \cite{Longstaff} introduced a least square regression method (LSM) to approximate the prices of American options. Much research has been done subsequently to give the analysis and the convergence of the LSM.\\ 
Lattice-type models for pricing options were implemented in many works (\cite{boyle1988lattice}; \cite{boyle1989numerical}). Various choices for tree parameters lead to different existing binomial models. The first proposed formulation of the binomial tree in financial mathematics was the research by Cox, Ross, and Rubinstein \cite{Cox1979}, providing a simplified discrete approach to option pricing on one asset using the recombining binomial trees based on Pascal's triangle. The authors in this article proposed to take $u=1/d$, where u, d are upward and downward movements. The work is still very popular today. One benefit of the model is that it exhibits a unique risk-neutral probability. Sierag and Hanzon \cite{Hanzon2018} extended this to multi-asset option pricing using recombining multinomial trees based on Pascal's simplex. The second popular model was the research by Jarrow and Rudd \cite{jarrow1983option}, also known as the equal-probability model.  One drawback is that the choice of equal probability seems to be unrealistic. Another drawback is that the model is no longer a risk-neutral model although it matches the risk-neutral continuous-time model in the limit for the time-step length going to zero.
Authors in (\cite{Cox1979}, \cite{jarrow1983option}) matched the first two moments to the risk-neutral continuous model, leading to a system of two equations for three unknowns. Tian's approach \cite{tian1993modified} was to equate the third moment to handle the issue of this free
variable. Leisen and Reimer \cite{leisen1996binomial} presented the most different choice of the model parameters $u,d$ and $p$ among the approaches mentioned above. They devoted attention to improving the convergence rate and smoothness when approximating the Gaussian distribution.
\\
The main advantage of classical methods such as the Tree or Lattice method is its simplicity for implementation in option pricing. 
A higher tree depth is required to get good precision. As binomial trees are recombining they greatly reduce the number of nodes compared to general binary trees. As a result, the corresponding computational cost  is reduced.\\
Options and other financial derivatives form an important section of the financial markets. The majority of exchange-traded options on a single asset tend to be American style, while options on indexes are European. 
After the financial crisis of 2007-2008, there is a requirement for a significant change in financial modelling and risk management, which is reflected in the Basel Accords-Basel I, Basel II, and Basel III for the calculations of the required capital issued by the BIS (Bank of International Settlements). The financial regulators required banks to hold an amount of capital to capture the credit risk in portfolios.  The BIS published Basel III to work along with Basel II in response to the deficiencies in banking regulation in Basel I. Basel III presented a new measure, namely Credit Valuation Adjustment to capture Counterparty Credit Risk (CCR) \cite{basel2017basel}. As a result there is an increased interest in CVA. This is the motivation for us to develop one  algorithm for the  calculation of the Credit Value Adjustment (CVA) on an American option, using our MC-Tree technique.\\
If the exposure profile and the credit quality of the counterparty are positively or negatively  related, so-called right-way risk (RWR) or wrong-way risk (WWR) occurs. Otherwise, one speaks about unilateral CVA.\\
In this article, unilateral CVA will be computed for an American put option based on standard assumptions, using the MC-Tree method. 
\section{MC-Tree Method}
Various existing versions of binomial models use different choices for parameters. We aim to improve the accuracy and speed of the binomial method by applying Monte Carlo simulation on these parameters. The idea is to generate the tree directions and probabilities through a parameter. This parameter is drawn from a probability density, called the mixing density. A formula will be provided for the resulting compound density that is then generated by the tree. The goal is to find a mixing density such that the corresponding compound density is close to a Gaussian density,  working with the additive representation of the tree. It does not seem to be possible to find a mixing density for which the compound density is exactly Gaussian. We succeed in specifying a class of mixing densities for which the compound density is rational (and hence smooth). The standard Gaussian density can be approximated by rational densities as follows from the following well-known limit: 
 $$\lim_{N \rightarrow \infty}\frac{1}{(1+\frac{x^2/2}{N})^{N}} = e^{-\frac{x^2}{2}}.$$
Therefore, it is not unreasonable to construct rational compound densities to approximate the standard Gaussian density function.
It can be proved that the standard Gaussian distribution can not be obtained in a 1-step tree. The reason is that standard Gaussian can be obtained on the negative half-line, which means that the mixing density can be computed so that the compound density is equal to the standard Gaussian density on the negative half-line.  However, this requirement fully determines the mixing density, and this mixing density produces a compound density on the positive half-line that is not a standard Gaussian at all.\\
To get good approximations of the Gaussian distribution, we try to find mixing densities for which the entropy of the compound density is high. The motivation is that our tree construction forces it to have a distribution on the end-nodes at step N with zero mean and variance equal to N. This will also hold for the compound density of any MC-Tree. If $\tau$ denotes a parameter characterizing a 1-step tree and X the random variable resulting from the MC-Tree procedure at time step N, then 
\[
\mathbb{E}_{X \mid \tau}(X)=0, \phantom{x}
\mathbb{E}_{X \mid \tau}(X^2)= N.
\]
It implies that $$\mathbb{E}_{X}(X)=\mathbb{E}_{\tau}(\mathbb{E}_{X \mid \tau}(X \mid \tau))=0, \phantom{x} \mathbb{E}_{X}(X^2)=\mathbb{E}_{\tau}(\mathbb{E}_{X \mid \tau}(X^2 \mid \tau))=N.$$
It is well-known from the literature that the standard Gaussian distribution $\mathbb{N}(0,1)$ with the pdf $f(x)=\frac{1}{\sqrt{2\pi}}\exp^{-\frac{x^2}{2}}$ maximizes the entropy integral\\
$Ent(f):=-\int_{\infty}^{\infty} f(x)log(f(x))dx$, subject to the constraints\\
$\int_{-\infty}^{\infty} f(x)dx=1,\ \int_{-\infty}^{\infty} xf(x)dx=0,\ \int_{-\infty}^{\infty} x^2f(x)dx=1$ and more generally a Gaussian density with given mean and variance maximizes the entropy among all densities with that same mean and variance.

\subsection{Parameters}
The multiplicative binomial tree has the probability $p_1>0$ of moving "down" to $S^{n}e^d$ and $p_2>0$ of moving "up" to $S^{n}e^u$, where $p_1+p_2=1$ and $S^n>0.$ Here, $u> d$; this is actually all that is required, and  $u>0$ and $d<0$ is not required. So for $n=0,1,\dots,$ we have

$$
  S^{(n+1)\delta t} = \left\{ \begin{array}{ll}
S^{n \delta t}e^u  & \textrm{w.p. $p_2$}\\
 S^{n \delta t}e^d  & \textrm{w.p. $p_1$.}
\end{array} \right.
$$

Equivalently, we have the following additive tree.
$$
log(S^{(n+1)\delta t}) = \left\{ \begin{array}{ll}
log(S^{n})+u & \textrm{w.p. $p_2$}\\
log(S^{n})+d & \textrm{w.p. $p_1$.}
\end{array} \right.
$$
As the mean and variance can always be adapted using an affine transformation, we will first consider the case in which mean=0 and variance=1 for each time-step. This implies
$$p_2+p_1=1,$$
$$p_2u+p_1d=0,$$
$$p_2u^2+p_1d^2=1,$$
$$p_2>0,\ p_1>0.$$
 The family of all solutions can be parametrized by an angle $\theta,$ with $0<\theta<\frac{\pi}{2},$ as follows:
\begin{itemize}
    \item $p_1^{1/2}=cos(\theta),$
    \item $p_2^{1/2}=sin(\theta),$
    \item $u=\sqrt{\frac{p_1}{p_2}},$
    \item $d=-\sqrt{\frac{p_2}{p_1}}.$
\end{itemize} 
For a geometric interpretation of the angle $\theta$, we refer to Sierag and Hanzon \cite{Hanzon2018}.
Hence, we now consider  $\theta$ to be a random variable with distribution function $P_m$, supported on $(0,\pi/2)$, providing us  with a mixing distribution on the tree. We can summarize this schematically as follows:
$$\log(S^{N}) \mid \theta \sim Binomial,\phantom{x}
\theta \sim P_m.$$
Here by "Binomial" we mean the distribution of log$(S^N)$ at the final nodes of the tree.
Combining the binomial distribution on the log-asset-prices with the distribution on $\theta$ will result in a compound density for log$(S^N).$ We will say more about how the compound density can be calculated in the next section.\\
The proposed approach is now to compute an expected value of a payoff function defined on log$(S^N),$ 
with respect to the compound density by (1) assembling, say, M  drawings of the variable $\theta$ and (2) for each $\theta$ to "run the tree" to obtain an approximation to the expected value of the payoff, and (3) to compute the average and standard deviation of the tree-outcomes to obtain a Monte Carlo estimation of the expected payoff, as well as a confidence interval. Informally we refer to this procedure as "shaking the tree". In terms of Monte Carlo theory, this method falls under the category "variance reduction by conditioning" \cite{brandimarte2013numerical}. The idea is that because the tree outcomes will already be very close to the true value (especially for deeper trees), the Monte Carlo outcomes will be very accurate. 

\subsection{Preliminaries to derivation of the general compound density formula}
Let $N \in \mathbb{N}.$ Define $$x_{N,k}=-(N-k)\tau + k \frac{1}{\tau}, \ \tau > 0,\ \ k=0,1,2,\ldots,N.$$
Note that 
$$
Range \ of \ x_{N,k} = \left\{ \begin{array}{lll}
(-\infty,0) & \textrm{if $k=0$, }\\
\mathbb{R} & \textrm{if $k=1,2,\dots,N-1$,}\\
(0,\infty) & \textrm{if $k=N$}.
\end{array} \right.
$$\\
As the derivative  $x'_{N,k}(\tau)=-(N-k)-\frac{k}{\tau^2} <0 \ \forall \ \tau \in (0,\infty),$ it follows that $x_{N,k}$ is monotonically decreasing for each $N \in \mathbb{N}$ and $k \in \{0,1,\dots,N\}.$ Hence, $x_{N,k}(\tau)$ has an inverse function $\tau_k(x)$ with domain $(-\infty,0)$ if $k=0,\ \mathbb{R}$ if $k=1,\dots,N-1$ and $(0,\infty)$ if $k=N$; and range $(0,\infty)$ in all cases.\\
As can easily be verified, one has
$$\tau_k(x)=\frac{-x+\sqrt{x^2+4k(N-k)}}{2(N-k)} \ \mbox{if} \ k \neq 0, N,$$
$$\tau_0(x)=-\frac{x}{N},\ x \in (-\infty,0),$$
$$\tau_N(x)=\frac{N}{x}, \ x \in (0,\infty).$$
Note that $\tau_k(x)\tau_{N-k}(-x)=1$ holds for all x for which the left hand side is defined. For later reference, we also define $y_k(x)=\sqrt{x^2+4k(N-k)},\ k=1,2,\dots,N-1.$ The variable $y_k$ can also be expressed in terms of $\tau_k$, as follows.
$$y_k(x)=(N-k)\tau_k +k \frac{1}{\tau_k}=(N-k)\tau_k(x)+k\tau_{N-k}(-x).$$
Now, for an additive binomial tree with $d=-\tau$ with the probability $ \frac{1}{1+\tau^2}$ and $u=\frac{1}{\tau}$ with the probability $ \frac{\tau^2}{1+\tau^2},$ the probability distribution of the values at the nodes at the tree depth N can be expressed as: 
$$
X_{N \mid\tau} = \left\{ \begin{array}{lll}
x_{N,0}(\tau) & \textrm{with prob $g_0(\tau):=\frac{1}{(1+\tau^2)^N}$,}\\
x_{N,k}(\tau) & \textrm{with prob $g_k(\tau):=\binom{N}{k}\frac{(\tau^{2})^k}{(1+\tau^2)^N}, \ k=\overline{1,N-1}$,}\\
x_{N,N}(\tau) & \textrm{with prob $g_N(\tau):=\frac{(\tau^{2})^N}{(1+\tau^2)^N}$}.
\end{array} \right.
$$
Note that $X_{N \mid \tau}$ is the sum of N stochastically independent copies of the random variable $X_{1 \mid \tau}$, and hence it has mean 
$\mathbb{E}[X_{N \mid \tau}]=0$, and the variance at 
$\mathbb{E}[(X_{N \mid \tau})^2]=N.$ So $\frac{1}{\sqrt{N}}X_{N \mid \tau}$ is a random variabe with mean zero and variance one.\\
Application of the Central Limit Theorem tells us that the Cumulative Distribution Function (CDF) of $\frac{1}{\sqrt{N}}X_{N \mid \tau}$ converges to the CDF of a Standard Gaussian random variable for $N \rightarrow \infty$ (and $\tau > 0$ fixed). The CDF of $X_{N \mid \tau}$ can be described as follows.
$$F(x \mid \tau)=P(X_{N \mid \tau} \leq x \mid \tau)=$$$$g_0(\tau)\mathbbm{1}_{\{x \geq x_{N,0}(\tau) \ \& \ x <0 \} }+ \sum_{k=1}^{N-1}g_k(\tau) \mathbbm{1}_{\{x \geq x_{N,k}(\tau)\}}+g_N(\tau)\mathbbm{1}_{\{x \geq x_{N,N}(\tau) >0 \}}.$$
If we now consider the tree parameter $\tau$ as random with the probability density function $p_m(\tau)$ (we will call $p_m(\tau)$ the "mixing density") then the resulting compound CDF of X will be $Q(x):=\int_{0}^{\infty}F(x \mid t)p_m(t)dt.$
\subsection{General Compound Density Formula}
\begin{theorem}
The compound probability density function $q(x)=Q'(x)$ of X satisfies the following formula
$$q(x)=\sum_{k=0}^{N}C_k(x),$$
where:
\begin{itemize}
\item $C_0(x)=\frac{1}{(1+\tau_{0}^{2})^{N}}p_m(\tau_{0})\frac{1}{N}\mathbbm{1}_{ \{x \leq 0 \} },$\\
\item $C_k(x)=\binom{N}{k}\frac{(\tau_{k}^{2})^k}{(1+\tau_k^2)^N}p_m(\tau_k)\frac{\tau_k}{y_k},\ k=\overline{1,N-1}.$\\
\item $C_N(x)=\frac{(\tau_{N}^{2})^N}{(1+\tau_N^2)^N}p_m(\tau_N)\frac{N}{x^2}\mathbbm{1}_{ \{x>0\} }.$
\end{itemize}
\end{theorem}
\begin{proof}
Fix N. As for each $k \in \{0,\dots,N\}$, the function $x_{N,k}(\tau)$ is monotonically decreasing with inverse $\tau_k(x)$, we can write
$$ Q(x)=\int_{0}^{\infty}F(x \mid \tau) p_m(\tau) d\tau =\mathbbm{1}_{\{x<0\}}\ \int_{\{\tau \geq \tau_0(x) \& \ x<0\}}g_0(\tau)p_m(\tau)d\tau+$$ 
$$\sum_{k=1}^{N-1}\int_{\{\tau \geq \tau_k(x)\}}g_k(\tau)p_m(\tau)d\tau +\mathbbm{1}_{\{x >0\}}\int_{\{ \tau \geq \tau_N(x) \ \& \ x>0\}}g_N(\tau)p_m(\tau)d\tau.$$
For $k=1,\dots,N-1$, we have 
$$1=\frac{d}{dx}x_{N,k}(\tau_k(x))=\{-(N-k)-\frac{k}{\tau_k^2}\}\tau'_k=-((N-k)\tau_k+k\frac{1}{\tau_k})\frac{\tau'_k}{\tau_k}=
-\frac{y_k}{\tau_k}\tau'_k.$$ 
Therefore, $$\tau'_k(x)=-\frac{\tau_k}{y_k},\ k=1,\dots,N-1.$$
Recall $\tau_0(x)=-\frac{x}{N} \Rightarrow \ \ \tau_0'(x)=-\frac{1}{N},\ x < 0,$ and $\tau_N(x)=\frac{N}{x} \Rightarrow \tau'_N(x)=-\frac{N}{x^2},\ x > 0.$
Now taking the derivative of $Q(x)$, we obtain
$$q(x)=\mathbbm{1}_{\{x <0\}}g_0(\tau_0(x))p_m(\tau_0(x))\frac{1}{N}+\sum_{k=1}^{N-1}g_k(\tau_k(x))p_m(\tau_k(x))\frac{\tau_k(x)}{y_k(x)}+$$
$$\mathbbm{1}_{\{x >0\}}g_N(\tau_N(x))p_m(\tau_N(x))\frac{N}{x^2}=C_0(x)+\sum_{k=1}^{N-1}C_k(x)+C_N(x).$$
\end{proof}
\subsection{Mixing Density}
As is well-known, due to the convexity of the function $g(y)=y \log(y),$ if $f(x)$ is a pdf on $\mathbf{R}$ then $\frac{f(x)+f(-x)}{2}$ has entropy at least as high as $f(x).$ 
As we are looking for compound densities with high entropy, the consequence of this is that we can restrict our search to mixing densities that produce an even compound density function. In terms of the mixing distribution, this translates into considering  mixing densities which are invariant under a permutation of the two axes in the binomial tree. So we use mixing probabilities on the recombining binomial tree such that symmetric paths have the same probability. 
A relatively simple class of mixing densities satisfying this invariance is given, in terms of the parameter $p_1,$
by
 $$\frac{1}{2}c_m(p_1^{1/2}p_2^{1/2})^{m-2}dp_1,\phantom{x} m \in \mathbb{N},$$
where $p_1>0,\ p_2>0,\ p_1+p_2=1$ and $c_m$ is a normalization constant.
Recalling $p_1=\cos^2(\theta),\ p_2=\sin^2(\theta), $ the transformation $\tau=\tan(\theta)$ leads to $p_1=\frac{1}{1+\tau^2},\ p_2=\frac{\tau^2}{1+\tau^2}$, so
 $\frac{1}{2}c_{m}(p_1^{\frac{1}{2}}p_2^{\frac{1}{2}})^{m-2}dp_1 
 =\frac{1}{2}c_{m}(\frac{1}{1+\tau^2})^{(m-2)/2}(\frac{\tau^2}{1+\tau^2})^{(m-2)/2}
  \lvert d(\frac{1}{1+\tau^2})\rvert =c_m \frac{\tau^{m-1}}{(1+\tau^2)^m}d\tau,$
where the constant $c_m$ is given by
$$c_m=\frac{1}{\int_0^{\infty}\frac{\tau^{m-1}}{(1+\tau^2)^m}d\tau}.$$
In terms of the parameter $\theta$ we obtain a third representation of these mixing densities:
$\frac{1}{2}c_{m}(p_1^{\frac{1}{2}}p_2^{\frac{1}{2}})^{m-2}dp_1
 =c_{m} (\cos(\theta)\sin(\theta))^{m-1}d\theta.$
The idea is now to apply the MC technique and draw $\tau$ from this  probability distribution on $(0,\infty)$. We will also  make use of the transformation $\tau=tan(\theta) \Longleftrightarrow \theta=arctan(\tau),\ \theta \in (0,\pi/2)$ regularly. Drawing $\tau$ can then be replaced by drawing $\theta$ and using $\tau=tan(\theta)$.
\subsection{Monte Carlo Drawing}
In order to carry out the Monte Carlo simulations we need to be able to draw independent samples from the mixing distribution. A general technique in case the cumulative distribution function (CDF) is available is to draw random samples from the uniform distribution on the interval $[0,1]$ and to use the inverse function of the CDF to obtain the desired samples. Note that as our mixing densities are everywhere positive that the inverse of its CDF exists, and given any drawing from the uniform distribution, the corresponding sample from the mixing distribution can be found, for instance, by a bisection method.
Therefore what remains is to find the CDF of our mixing distributions. One way to do that is to work out the CDF of the mixing density in terms of the angle $\theta.$ This can be done as follows:\\
Note that
$$p_m(\theta)=c_m\ cos(\theta)^{m-1}\sin(\theta)^{m-1}=
c_m (\frac{e^{i\theta}+e^{-i\theta}}{2})^{m-1}(\frac{e^{i\theta}-e^{-i\theta}}{2i})^{m-1}=
$$
$$c_m2^{-2(m-1)} \ Re[(-i)^{m-1}(e^{i2\theta}-e^{-i2\theta})^{m-1}]=$$
$$c_m2^{-2(m-1)} \ Re[(-i)^{m-1}e^{-i2\theta (m-1)}((e^{i4\theta}-1)^{m-1}]=$$
$$c_m2^{-2(m-1)} \ Re[(-i)^{m-1}e^{-i2\theta (m-1)}\sum_{s=0}^{m-1}\binom{m-1}{s}e^{i4s\theta}(-1)^{m-1-s}]=$$
$$c_m2^{-2(m-1)} \ Re[(-i)^{m-1}\sum_{s=0}^{m-1}\binom{m-1}{s}e^{i(4s-2(m-1))\theta}(-1)^{m-1-s}].$$
\begin{itemize}
\item In case m is odd, this has the following primitive function
$$c_m2^{-2(m-1)} \ Re[(-i)^{m-1}\sum_{s=0,\ s \neq \frac{m-1}{2}}^{m-1}\binom{m-1}{s}\frac{(-1)^{m-1-s}}{i(4s-2(m-1))}e^{i(4s-2(m-1))\theta}+$$$$(-i)^{m-1}\binom{m-1}{(m-1)/2}\theta (-1)^{(m-1)/2}+\Tilde{C}_m]=$$
$$c_m2^{(-2m+1)} \ Re[(-i)^{m}\sum_{s=0,\ s \neq \frac{m-1}{2}}^{m-1}\binom{m-1}{s}\frac{(-1)^{m-1-s}}{(2s-(m-1))}e^{i(4s-2(m-1))\theta}+$$$$(-1)^{(m-1)/2}\binom{m-1}{(m-1)/2}\theta (-1)^{(m-1)/2}+\Tilde{C}_m]=$$
$$c_m2^{(-2m+1)}[(-1)^{(m-1)/2}\sum_{s=0}^{m-1}\binom{m-1}{s}\frac{(-1)^{m-1-s}}{(2s-(m-1))}\sin((4s-2(m-1))\theta)+$$
$$\binom{m-1}{(m-1)/2}\theta +\Tilde{C}_m],$$
where $\Tilde{C}_m, \ m=1,2,...$ are real integration constants. As the CDF $F_m(\theta)$ has its support on $(0,\frac{\pi}{2})$, we have $F_m(0)=0.$ In case m is odd, $\Tilde{C}_m=0.$\\
We implies 
$$c_m=\frac{2^{2m-1}}{\binom{m-1}{(m-1)/2}\frac{\pi}{2}}.$$
\item In case m is even, we can perform similar calculations, as follows:\\ 
$p_m(\theta)$ has the following primitive function
$$c_m2^{-(m-1)} \ Re[(-i)^{m-1}\sum_{s=0}^{m-1}\binom{m-1}{s}\frac{(-1)^{m-1-s}}{i(4s-2(m-1))}e^{i(4s-2(m-1))\theta}+\Tilde{C}_m]=$$
$$c_m2^{(-2m+1)} \ Re[(-i)^{m}\sum_{s=0}^{m-1}\binom{m-1}{s}\frac{(-1)^{m-1-s}}{(2s-(m-1))}e^{i(4s-2(m-1))\theta}+\Tilde{C}_m]=$$
$$c_m2^{(-2m+1)}(-1)^{(m/2)}\sum_{s=0}^{m-1}\binom{m-1}{s}\frac{(-1)^{m-1-s}}{(2s-(m-1))}\cos((4s-2(m-1))\theta)+\Tilde{C}_m.$$
 $$\Tilde{C}_m=c_m2^{(-2m+1)}(-1)^{(m/2)+1}\sum_{s=0}^{m-1}\binom{m-1}{s}\frac{1}{(2s-(m-1))}(-1)^{m-1-s}.$$
 \end{itemize}
\subsection{The Particular Compound Density}
\begin{theorem}  Let $m$ be odd and let the mixing density be given by $p(\tau)=c_m\frac{\tau^{m-1}}{(1+\tau^2)^m},$ where $c_m$ is the normalizing constant (as before).
The compound density takes the form
$q(x)=c_m \ \frac{A_m(x)}{(x^2+N^2)^{N+m}},$ where $A_m(x)$ is a polynomial with rational coefficients and degree at most $2(N+m-1)$.
\end{theorem}
\begin{proof}
We consider two following cases.
\begin{itemize}
\item N even:
$$q(x)=(C_0+C_N)(x)+C_{N/2}(x)+\sum_{k=1}^{\frac{N}{2}-1}(C_k+C_{N-k})(x).$$
\item N odd:
$$q(x)=(C_0+C_N)(x)+\sum_{k=1}^{\frac{N-1}{2}}(C_k+C_{N-k})(x).$$
\end{itemize}
We have
$$C_0=\frac{1}{(1+\tau_0^2)^N}p_m(\tau_0)\frac{1}{N}\mathbbm{1}_{\{x\leq 0\}}=c_m\frac{\tau_0^{m-1}}{(1+\tau_0^2)^{N+m}}\frac{1}{N}\mathbbm{1}_{\{x\leq 0\}}=$$
$$c_m (-1)^{m-1}\frac{x^{m-1}N^{(2N+m)}}{(x^2+N^2)^{N+m}}\mathbbm{1}_{\{x\leq 0\}}=c_m \frac{x^{m-1}N^{(2N+m)}}{(x^2+N^2)^{N+m}}\mathbbm{1}_{\{x\leq 0\}}.$$ Here we have used $(-1)^{m-1}=1$ as $m$ is odd.\\
$$C_N=\frac{(\tau_N^2)^N}{(1+\tau_N^2)^N}p_m(\tau_N)\frac{N}{x^2}\mathbbm{1}_{\{x > 0\}}=c_m\frac{\tau_N^{2N+m-1}}{(1+\tau_N^2)^{N+m}}\frac{N}{x^2}\mathbbm{1}_{\{x > 0\}}=$$
$$c_m \frac{x^{m-1}N^{(2N+m)}}{(x^2+N^2)^{N+m}}\mathbbm{1}_{\{x > 0\}.}$$
It follows that the first term $(C_0+C_N)(x)$ can be written explicitly as follows for all real values of $x$:
$$(C_0+C_N)(x)=c_m\frac{N^{2N+m}x^{m-1}}{(x^2+N^2)^{N+m}}=c_m\frac{A_m(x,0)}{(x^2+N^2)^{N+m}},$$
where $A_m(x,0):=N^{2N+m}x^{m-1}.$ Recall that (m-1) is even, so $(C_0+C_N)(x)$ is even w.r.t x.\\
Now we consider the case $1 \leq k\leq N.$
Observe that $$(\frac{-x+y_k}{2(N-k)})(\frac{x+y_k}{2k})=\frac{-x^2+y_k^2}{4(N-k)k}=\frac{-x^2+x^2+4k(N-k)}{4k(N-k)}=1.$$
Hence, as $\tau_k=\frac{-x+y_k}{2(N-k)}$ then $\tau_k^{-1}=\frac{x+y_k}{2k}$ and $\tau_{N-k}^{-1}=\frac{x+y_k}{2(N-k)}.$
We rewrite $C_{N-k}$ in terms of $\tau_{N-k}^{-1}$ as follows:
\begin{eqnarray*}
\lefteqn{C_{N-k}=\binom{N}{N-k}\frac{(\tau^2_{N-k})^{N-k}}{(1+\tau_{N-k}^2)^N}p_m(\tau_{N-k})\frac{\tau_{N-k}}{y_{N-k}}}\\
& & =c_m\binom{N}{N-k}\frac{(\tau^2_{N-k})^{N-k}}{(1+\tau_{N-k}^2)^N}\frac{\tau_{N-k}^{m-1}}{(1+\tau_{N-k}^2)^m}\frac{\tau_{N-k}}{y_{N-k}}\\
&&=c_m\binom{N}{N-k}\frac{(\tau_{N-k}^2)^{N-k+\frac{m-1}{2}+1}\tau_{N-k}^{-1}}{(1+\tau_{N-k}^2)^{N+m}y_{N-k}}\\
& & =c_m\binom{N}{N-k}\frac{\frac{1}{(\tau_{N-k}^{-2})^{N-k+\frac{m-1}{2}+1}}\tau_{N-k}^{-1}}{(1+\frac{1}{\tau_{N-k}^{-2}})^{N+m}y_{N-k}}\\
&&=c_m\binom{N}{N-k}\frac{(\tau^{-2}_{N-k})^{k+\frac{m-1}{2}}\tau_{N-k}^{-1}}{(1+\tau_{N-k}^{-2})^{N+m}y_{N-k}}.
\end{eqnarray*}
Notice that $y_{N-k}=y_k$ and $\binom{N}{N-k}=\binom{N}{k}.$
The term $C_k+C_{N-k}$ can be defined w.r.t x and $y_k$ as follows:
$$(C_k+C_{N-k})(x,y_k)=c_m\binom{N}{k}\frac{(\tau^{2}_{k})^{k+\frac{m-1}{2}}\tau_{k}}{(1+\tau_{k}^{2})^{N+m}y_{k}}+c_m\binom{N}{k}\frac{(\tau^{-2}_{N-k})^{k+\frac{m-1}{2}}\tau_{N-k}^{-1}}{(1+\tau_{N-k}^{-2})^{N+m}y_{N-k}}$$
$$=c_m\binom{N}{k}\frac{1}{2(N-k)}[\frac{(\tau_k^2)^{k+\frac{m-1}{2}}}{(1+\tau_k^2)^{N+m}}(\frac{-x}{y_k}+1)+\frac{(\tau_{N-k}^{-2})^{k+\frac{m-1}{2}}}{(1+\tau_{N-k}^{-2})^{N+m}}(\frac{x}{y_k}+1)]$$
$$=c_m\binom{N}{k}\frac{1}{2(N-k)}[\frac{((\frac{-x+y_k}{2(N-k)})^2)^{k+\frac{m-1}{2}}}{(1+(\frac{-x+y_k}{2(N-k)})^2)^{N+m}}(\frac{-x}{y_k}+1)+\frac{((\frac{x+y_k}{2(N-k)})^{2})^{k+\frac{m-1}{2}}}{(1+(\frac{x+y_k}{2(N-k)})^{2})^{N+m}}(\frac{x}{y_k}+1)]$$
$$=c_m\binom{N}{k}\frac{(2(N-k))^{2N-2k+m}N(x,y_k)}{D(x,y_k)},$$
where $N(x,y_k)=(4(N-k)^2+(-x+y_k)^2)^{N+m}(x+y_k)^{2k+m}+(4(N-k)^2+(x+y_k)^2)^{N+m}(-x+y_k)^{2k+m}.$
$$D(x,y_k)=((4(N-k)^2+x^2+y_k^2)^2-4x^2y_k^2)^{N+m}y_k.$$
Observe that $N(x,y_k)=-N(x,-y_k)$ and $D(x,y_k)=-D(x,-y_k)$, so $\frac{N(x,y_k)}{D(x,y_k)}=\frac{N(x,-y_k)}{D(x,-y_k)}$.
Hence, $(C_k+C_{N-k})(x,y_k)$ is even in $y_k$.
We can see that $$N(x,0)=(4(N-k)^2+(x)^2)^{N+m}(x)^{2k+m}-(4(N-k)^2+(x)^2)^{N+m}(x)^{2k+m}=0$$
Hence, we infer that $N(x,y_k)$ is divisible by $y_k,$ so both numerator and denominator are divisible by $y_k$. 
We have $D(x,y_k)/y_k$ and $C_k+C_{N-k}$ both are even in $y_k$, so $N(x,y_k)/y_k$ is even in $y_k$. This implies that we can express the new numerator $N(x,y_k)/y_k$ and the new denominator $D(x,y_k)/y_k$ both as polynomial in terms of powers of $x$ and  powers of $y_k^2$. Replacing $y_k^2=x^2+4k(N-k)$, we can conclude that $(C_k+C_{N-k})(x)$ is rational in x. In a similar way we can see that  $(C_k+C_{N-k})(x)$ is an even function of $x.$ \\
Observe that 
$$((4(N-k)^2+x^2+y_k^2)^2-4x^2y_k^2)=$$
$$(4(N-k)^2+x^2+x^2+4k(N-k)^2)^2-4x^2(x^2+4k(N-k))=$$
$$4\{(2(N-k)^2+x^2+2k(N-k))^2-x^4-4k(N-k)x^2\}=$$
$$4\{(2N(N-k)+x^2)^2-x^4-4k(N-k)x^2\}=$$
$$4\{x^4+4N(N-k)x^2+4N^2(N-k)^2-x^4-4k(N-k)x^2\}=$$
$$16(N-k)\{(N-k)x^2+N^2(N-k)\}=16(N-k)^2\{x^2+N^2\}.$$
Hence, $$(C_k+C_{N-k})(x,y_k)=c_m \frac{\Tilde{N}(x,y_k)}{\frac{2^{2N+2k+3m}(N-k)^{2k+m}}{\binom{N}{k}}(x^2+N^2)^{N+m}},$$ where 
$$\Tilde{N}(x,y_k)=\frac{N(x,y_k)}{y_k},\ d_k:= \frac{2^{2N+2k+3m}(N-k)^{2k+m}}{\binom{N}{k}},\ A_m(x,k):=\frac{\Tilde{N}(x,y_k)}{d_k}.$$
Therefore, the term $(C_k+C_{N-k})(x)$ can hence be written as
$$(C_k+C_{N-k})(x)=c_m\frac{A_m(x,k)}{(x^2+N^2)^{N+m}},$$ where $A_m(x,k)$ is a polynomial with rational coefficients.\\ Note that in case $k=N/2$ this formula implies that $C_k(x)=\frac{1}{2}c_m\frac{A_m(x,k)}{(x^2+N^2)^{N+m}}.$ The compound density $ q(x)$ is an even rational function with common denominator $(x^2+N^2)^{N+m}$ because it is the sum
of even rational functions with the same denominator. The numerator of $q/c_m$ is the sum of polynomials with rational coefficients, hence is a polynomial with rational coefficients. Notice that as $q$ has integral one over the real line and each of the $(C_k+C_{N-k})(x)$ functions is non-negative, each such function is integrable and hence its codegree must be at least 2. The same argument holds for $q$ itself and so the numerator degree of $q$ will be less than or equal to $2(N+m-1).$
\end{proof}
To compute $A_m(x,k)$, one could use algebraic manipulation with Euler substitution to eliminate all the occurrences of square roots. Alternatively, one could compute $A_m(x,k)$ using a Lagrange interpolation technique that we will now explain. 
We need to take $(N+m)$ interpolation  points $x_k^i, i=1,2,\ldots,N+m$ to approximate $A_m(x,k)$ because the degree of numerator $A_m(x,k)$ is at most 2(N+m-1) and $A_m(x,k)$ is even.  
Here k is fixed for each term $(C_k+C_{N-k})(x).$
We can calculate the values of $A_m(x_k^i,k)$ by noting that
$$A_m(x_k^i,k)=\frac{1}{c_m}(C_k+C_{N-k})(x_k^i)((x_k^i)^2+N^2)^{N+m}.$$
 By applying Lagrange interpolation method, we can obtain
 {\tiny{
 \begin{displaymath}
 \left(
 \begin{array}{c}
A_m(x_k^1,k)  \\
A_m(x_k^2,k)  \\
\vdots\\
A_m(x_k^{N+m},k)
\end{array} \right)
 =
\left(
\begin{array}{ccccc}
1 & (x_k^1)^2&(x_k^1)^4 & \ldots&(x_k^1)^{2(N+m-1)} \\
1 & (x_k^2)^2&(x_k^2)^4 &\ldots& (x_k^2)^{2(N+m-1)} \\
\vdots & \vdots & \vdots&\vdots&\vdots\\
1&(x_k^{(N+m)})^2&(x_k^{(N+m)})^4&\ldots & (x_k^{(N+m)})^{2(N+m-1)}
\end{array} \right)
 \left(
 \begin{array}{c}
a_0  \\
a_2  \\
\vdots\\
a_{2(N+m-1)}
\end{array} \right)
\end{displaymath}}}
The matrix is a (N+m) x (N+m) Vandermonde
matrix of interpolation points. It is known to be non-singular as the interpolation points will be distinct. The Lagrange matrix is the known inverse matrix of this Vandermonde matrix, so we can obtain the solution by using the Lagrange coefficients explicitly. Alternatively we can solve this linear system of equations directly by standard methods.\\
Recall $A_m(x,k)=\Tilde{N}(x,y_k)/d_k$, where $\Tilde{N}(x,y)=\frac{N(x,y)}{y}$ is a known two-variable polynomial in x and y with integer coefficients and $d_k$ is a known integer. To be able to get the rational coefficients of $A_m(x,k)$ exactly, we need to take the interpolation points such that {\em both} $x_k^i$ and $y_k(x_k^i)$ are rational! That is indeed possible as we will now show.
Our approach will be based on an Euler substitution (known from the theory of integration).\\
Let $z_k:=-x+y_k=-x+\sqrt{x^2+4k(N-k)}$ then $z_k+x=y_k $. It implies that
$$z_k^2+x^2+2z_k x=x^2+4k(N-k) \Leftrightarrow 2z_k x=4k(N-k)-z_k^2 \Leftrightarrow x=\frac{2k(N-k)}{z_k}-\frac{1}{2}z_k. $$
It follows that $y_k=x+z_k=\frac{2k(N-k)}{z_k}+\frac{1}{2}z_k.$\\
Note that if we choose $z_k$ rational and non-zero ($z_k \in \mathbb{Q} \setminus \{0\}$) then both x and $y_k$ will be rational. Furthermore, for any $\hat{x} \in \mathbb{R}$ one can calculate $\hat{z}_k=-\hat{x} +\sqrt{\hat{x}^2+4k(N-k)} > 0$ and take a positive rational number $z_k$ arbitrarily close to $\hat{z}_k$ and compute the corresponding rational values of x and $y_k$. By taking $z_k$ sufficiently close to $\hat{z}_k$, the corresponding values x and $y_k$ will be as close as is desired to $\hat{x}$ and $\hat{y_k}$. (Warning: care must be taken for cases in which $\hat{z}_k$ is close to zero).\\
The Lagrangian interpolation method now requires us to solve a Vandermonde-type linear system of (N+m) equations with only {\em rational} coefficients. The solution will be a vector of rational numbers in $\mathbb{Q}^{N+m}$. How to solve such systems in case N+m is large is an active area of research in which considerable advances have been made \cite{steffy2011exact}.
In our project, we have relied on existing routines in symbolic computation.\\
\textbf{Remark 1.\\}
A random variable X has the mean at zero and the variance at N with the PDF q(x). Then $Z:=\frac{X}{\sqrt{N}}$ has the mean at zero and  the variance at 1 with the PDF $\sqrt{N}q(z\sqrt{N}).$\\
\textbf{Remark 2.\\}
MC-Tree allows to generate probability parameters and other parameters randomly with a well-known chosen probability distribution.
MC-Tree holds all benefits of both the MC method and the tree method. Also, one advantage of the MC-Tree approach is that it allows us to work with the confidence interval of the MC simulations. If we use only the tree method, we can not have the confidence interval and its benefit. The confidence interval depends on the number of simulations.\\
\textbf{Remark 3.\\}
For numerical implementation in section 5, we only search among odd values of m  as these are the only ones giving a rational compound density. Based on our numerical results on entropy, but also on the Kullback-Leibler (KL) Divergence, as well as $L^1$ distance, the best choice for the integer $m$ in our class of mixing densities is at $m=9$. 
\section{The Correction Techniques}
\subsection{Usage of a Bias-Correction}
To compute a European option, a price process $S_t$ is modeled by the geometric Brownian motion under its associated "risk-neutral" measure $\mathrm{Q}$ $$dS_t=r S_tdt+\sigma S_td\overline{W}_t,$$
where $r \in \mathbb{R}$ is the (constant) interest rate, $\sigma \in \mathbb{R}$ represents the diffusion coefficient, and $\overline{W}$ is a standard Brownian motion process under $\mathbb{Q}$. Using Itô's lemma with f(S) = log(S) gives a classic result, in which the process log(S) follows the normal distribution $\mathbb{N}((r - \sigma^2/2)T,\sigma^2T)$ on any interval of length T. This process can be approximated by a binomial tree.\\
We consider for instance the payoff functions $\pi(X)=Max(e^X-K,0)$ for a call option or $\pi(X)=Max(K-e^X,0)$ for a put option, where X=log(S).\\
For option pricing one typically uses the multiplicative tree (as mentioned before). Using this a multiplicative upward move, $u_1$, and a multiplicative downward move, $d_1$, are defined as
$$ u_1=\exp\{u\sigma\sqrt{\delta t}+(r-\sigma^2/2)\delta t \},\ d_1=\exp\{d\sigma\sqrt{\delta t}+(r-\sigma^2/2)\delta t\}.$$ 
Here $\delta t>0$ denotes the time-step length and translation and scaling have been applied to introduce the volatility parameter $\sigma$ and the drift term $r.$ 
Using bias-corrected directions (see \cite{Hanzon2018}) gives
$$\Tilde{u}_1:=u_1e^{-\lambda\delta t},\ \Tilde{d}_1:=d_1e^{-\lambda\delta t},$$ where the real number $\lambda$ is solved from $p\Tilde{u}_1+(1-p)\Tilde{d}_1=e^{r\delta t}.$ We obtain
$$\lambda=\frac{log(pu_1+(1-p)d_1)}{\delta t}-r.$$
This correction amounts to replacing $\hat{\mu}=r-\frac{1}{2}\sigma^2$ by $\hat{\mu}=r-\frac{1}{2}\sigma^2+\lambda.$ 
The resulting tree model is complete and free of arbitrage and has the probability given by $p_1$ and $p_2$ as the risk-neutral probability in the tree (see \cite{Hanzon2018}).
\subsection{Usage of a Distribution Correction Factor}
Consider the  compound density now with the appropriate variance and mean (obtained by scaling and translation as in the previous section). With some abuse of notation we will denote this again by $q$. Suppose we use a mixing density, say $p_m$ with $m=9$ for instance. Then the compound density $q$ will be close to Gaussian, in the sense that it has high entropy, especially for deeper trees, but it is not exactly equal to the corresponding Gaussian density with same mean and variance. To compensate for that, one can employ a {\em distribution correction factor}. This technique is known from the Monte Carlo method of importance sampling. The distribution correction factor, which we will denote by C(x), can be derived, in the context of option pricing, as follows:\\
 Let P denotes the price of an European option with payoff $\pi(X)$ at time T, then we have
 $$P=e^{-rT}\mathbb{E}_{Q}[\pi(X)]=e^{-rT}\int \pi(x)f(x)dx=e^{-rT}\int [\pi(x)\frac{f(x)}{q(x)}]q(x)dx $$
 $$=e^{-rT}\int [\pi(x)C(x)]q(x)dx=e^{-rT}\mathbb{E}_{q}[\pi(X)C(X)],$$
where q(x) is the compound density and we have the Gaussian density $f(x) \sim \mathrm{N}((r-\frac{1}{2}\sigma^2)T, \sigma^2T).$ 
In this way we get an {\em exact} Monte Carlo method for the pricing of European options that depend only on the asset price at expiry, in the sense that the compensated compound density is exactly Gaussian. 
\section{Unilateral CVA}
Modelling CVA is complicated because it consists of at least three components: Probability of Default (PD), Loss Given Default (LGD), and Exposure at Default (EAD). We can make some standard assumptions. Assume that LGD and discount factors are nonrandom. We assume the possible default event of the counterparty and the value of a netting set are uncorrelated, which is an assumption of unilateral CVA. 
We will use an intensity default model  \cite{brigo2006interest} to calculate the default probability of a counterparty. The exposure calculations depend highly on the complexity of the derivatives in the portfolio. Modelling CVA of an American option is a challenge due to the complexity of CVA calculations and the characteristics of the American option. 
 We present an algorithm for the calculation of the expected exposure in the formula of the unilateral CVA for the American put option using MC-Tree in the next section. Our method of calculations for the CVA of an American option is not known in the literature, to the best knowledge of the authors.
\subsection{Tree Approach for CVA}
We develop the algorithm to calculate CVA on an American put option, using the MC-tree method, as follows:\\
\textbf{Step 1:} Run the tree backward to compute the American option value at each node. Label each node either C for continuation or N for no continuation. Let \begin{displaymath}
\mathbbm{1}_{C} = \left\{ \begin{array}{ll}
1 & \textrm{if node label = "C",}\\
0 & \textrm{if node label = "N".}
\end{array} \right.
\end{displaymath}
\textbf{Step 2:} Run the tree forward to compute the probabilities of the American option reaching a given node.\\
\begin{figure}[ht!]
\includegraphics[height=2.2cm,width=12.21cm]{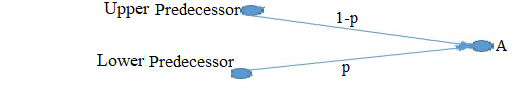}
\end{figure}\\
$$P(A)= p*P(Lower\quad Predecessor)\mathbbm{1}_{C}(Lower\quad Predecessor)+$$
$$(1-p)*P(Upper\quad Predecessor)\mathbbm{1}_{C}(Upper\quad Predecessor).$$
\textbf{Step 3:} Run the tree backward to compute the expected exposure at time step in the tree using the probabilities in step 2.
\subsection{MC-Tree Approach for CVA}
The tree approach for CVA will produce a CVA value for any given tree. The CVA value from the MC-Tree approach is the mean of all CVA values of all trees. 
\section{Numerical Results}
\subsection{Pricing European Options}
\begin{figure}[ht!]
\begin{center}
\includegraphics[height=8.75cm,width=11.36cm]{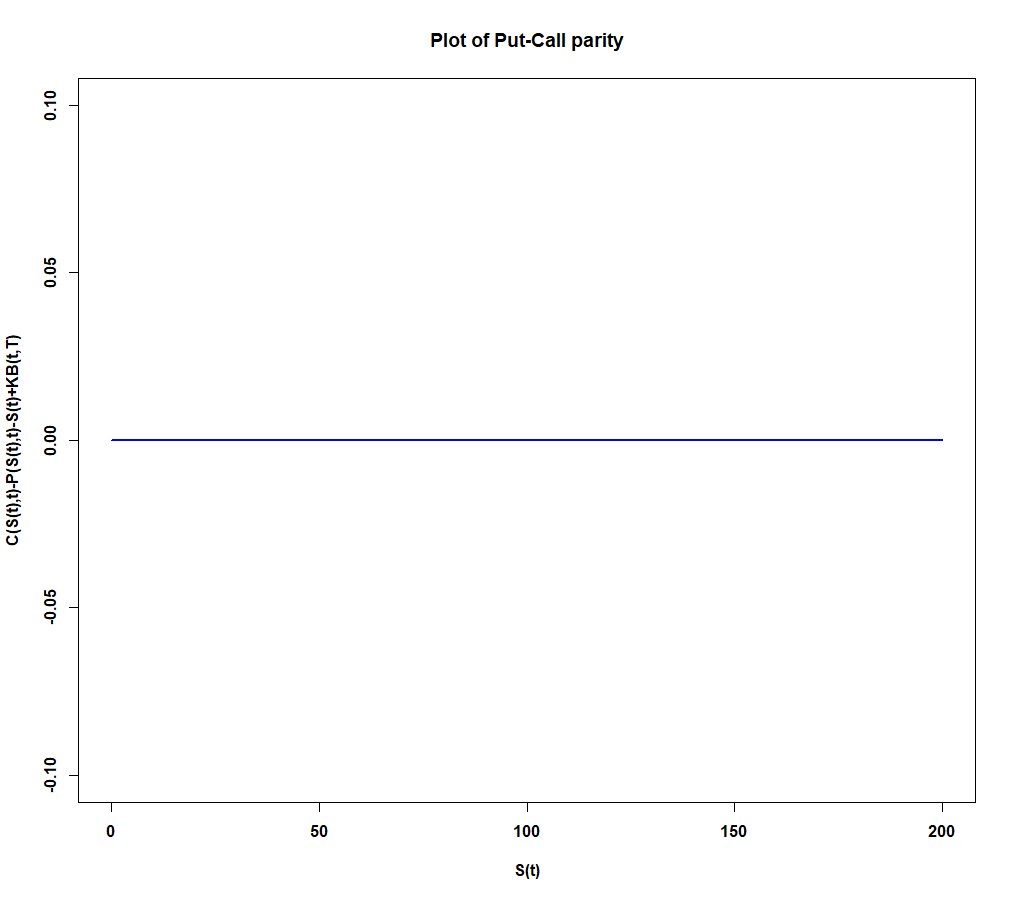}
\caption{Verification of Put-Call Parity}\label{PCParity}
\end{center}
\end{figure}

We will present the numerical results of some experiments to pricing the European option and compare MC-Tree with the usage of the bias-correction and the distribution correction factor with the Monte Carlo (MC) method and popular binomial tree models. We can obtain the analytical solution from the well-known Black-Scholes model. The error is the difference between the model value and the analytical solution.\\
All experiments were conducted on the machine I5-10210U, 8GB memory, R i386 3.5.1. The following parameters are used through all numerical experiments.
\begin{itemize}
    \item Strike price $K=95$.
    \item Expiration time $T=1$.
    \item Risk-free rate $r=0.03$.
    \item Volatility $\sigma=0.2$.
\end{itemize}
It is verified that the put-call parity is hold for the MC-Tree with the bias-correction, as shown in Figure \ref{PCParity}.
 \subsubsection{Comparison to MC Method }
Table \ref{comparemctrelmcallopt} and table \ref{comparemctrelmcputopt} show that MC-Tree is more accurate than MC method. The usage of the distribution correction factor in option pricing improves the accuracy dramatically, resulting in the exact analytical solution at the tree depth N=50 or 100, and the number of simulations M=100000.
\begin{table}
\centering
\begin{tabular}{|l|l|l|l|l|l|l|}
\hline
S & N&Method&Mean&SD&CI&AS\\ 
 \hline
100&50 &Corr &12.1798   &0.025 &(12.17965, 12.17995) &12.1797\\
&50 &Bias &12.1905  &0.0279 &(12.1903, 12.1907) &\\
& 100&Corr &12.1797 &0.0123 & (12.17962, 12.17978)&\\ 
&100 &Bias&12.1851  &0.0155 &(12.1850, 12.1852) &\\
& &MC&12.1867 &15.6215 &(12.0899, 12.2835) &\\
\hline
90&50 &Corr &6.2125 &0.071 & (6.2121, 6.2130)&6.2125\\
&50 &Bias &6.2230 &0.0596 &(6.2226, 6.2233) &\\
& 100&Corr &6.2125 &0.0463 & (6.212213, 6.212787)&\\ 
&100 &Bias&6.2177 &0.0401 &(6.2175, 6.2180) &\\
& &MC&6.2143 &11.1305 &(6.1453, 6.2833) &\\
\hline
\end{tabular}
\caption{Accuracy Comparison between MC-Tree and MC Method in Pricing European Call Option. Bias: MC-Tree with the usage of Bias-Correction; Corr: MC-Tree with the usage of the distribution correction factor. }\label{comparemctrelmcallopt}
\end{table}
 \begin{table}
\centering
\begin{tabular}{|l|l|l|l|l|l|l|}
\hline
S & N&Method&Mean&SD&CI&AS\\ 
 \hline
100&50 &Corr &4.3720 &0.0324 &(4.3718, 4.3722) &4.3720\\
&50 &Bias &4.3828 &0.0279 &(4.3827, 4.3830 ) &\\
& 100&Corr &4.3720 & 0.0185& (4.3719, 4.3721)

&\\ 
&100 &Bias&4.3774 &0.0155 & (4.3773, 4.3775 )&\\
& &MC& 4.4107&7.6584 &(4.3633, 4.4582) &\\
\hline
90&50 &Corr &8.4048 &0.0503 &(8.4045, 8.4051) &8.4048\\
&50 &Bias &8.4153 &0.0596 &(8.4149, 8.4157) &\\
& 100&Corr &8.4048 &0.0345 &(8.4046, 8.4050) &\\ 
&100 &Bias&8.4101 &0.0401 &(8.4098, 8.4103) &\\
& &MC&8.4352 &10.1748 &(8.3721, 8.4982) &\\
\hline
\end{tabular}
\caption{Accuracy Comparison between MC-Tree and MC Method in Pricing European Put Option. Bias: MC-Tree with the usage of Bias-Correction; Corr: MC-Tree with the usage of the distribution correction factor. }\label{comparemctrelmcputopt}
\end{table}
\\
Table \ref{Table12} shows the results from both methods for similar computation time.
\begin{table}[h]
\centering
\begin{tabular}{|l | l | l|}
\hline
 & MC-Tree&MC \\
  & with correction& \\
\hline
Option Price&12.1798 & 12.1800  \\
\hline
SD&0.025&15.6149\\
\hline
CI&(12.17965, 12.17995) &(12.17117,12.18883)\\
\hline
Error&5.7e-05&3e-04  \\
\hline
Computation Time (Seconds)&34.92103  & 35.50149  \\
\hline
M&100000& 12000000 \\
\hline
\end{tabular}
\caption{Error and Computation time of European call Option with similar running time at N=50.}\label{Table12}
\end{table}
It is evident from table \ref{Table12} that the MC-Tree model is still more accurate than the MC method, even with the similar computation time.
\subsubsection{Comparison to Binomial Models}
\begin{figure}[ht!]
\begin{center}
\includegraphics[height=15.75cm,width=11.36cm]{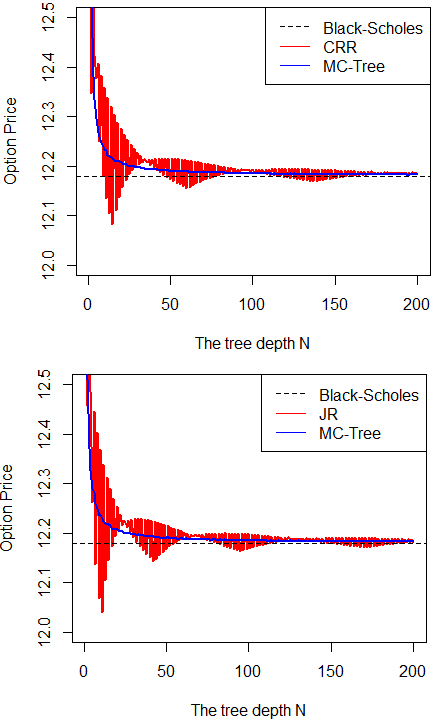}
\caption{Option prices v.s the tree depth N.}\label{Fig5}
\end{center}
\end{figure}

\begin{figure}[ht!]
\begin{center}
\includegraphics[height=6.75cm,width=10.36cm]{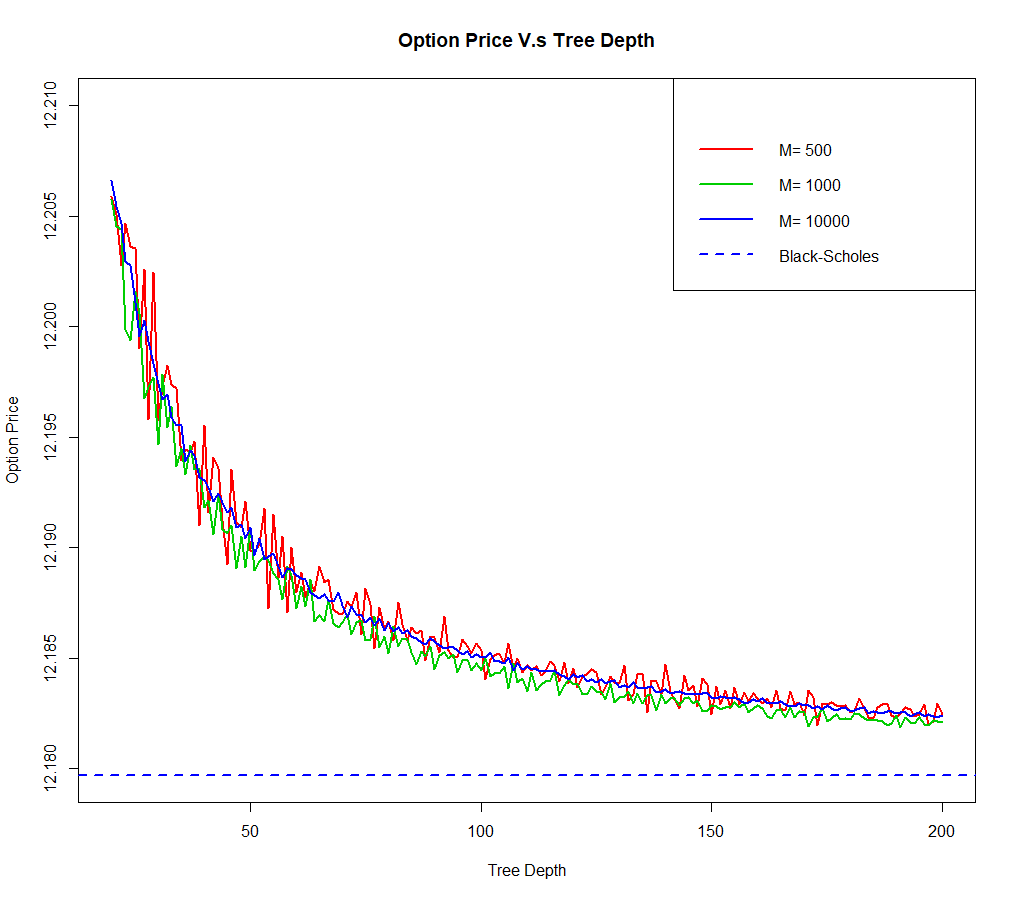}
\caption{European Call Prices Versus MC Drawings
M and the Tree Depth N.}\label{PlotversusNandM}
\end{center}
\end{figure}

Table \ref{mctreewithbinmodelscallopt} and table \ref{mctreewithbinmodelsputopt} show that MC-Tree with the usage of the distribution correction factor performs best. Option price from MC-Tree quicker converges to the analytical price when increasing the tree depth.
\begin{table}[h]
\centering
\begin{tabular}{|l|l|l|l|l|l|l|l|l|}
\hline
S & Method&
\multicolumn{3}{c|}{$N=50$}
&\multicolumn{3}{c|}{$N=100$}&
AS\\ 
 & &Mean&SD&CI & Mean&SD&CI  &  \\
\hline
100 &Corr &12.1798   &0.025 &(12.1797, &12.1797&0.0123&(12.1796, &12.1797\\
 & &  & &12.1800) &&&12.1798)&\\
  &Bias &12.1905  &0.0279 &(12.1903,  &12.1851 &0.0155&(12.1850, &\\
  & &  & & 12.1907) & && 12.1852)&\\
 &CRR  &12.1733&&&12.1923&& & \\  
 &JR  &12.1677&&&12.1984&& & \\ 
\hline
90 &Corr &6.2125 &0.071 &(6.2121,  &6.2125&0.0463&(6.2122, &6.2125\\
 & & & & 6.2130) &&& 6.2128)&\\
  &Bias &6.2230 &0.0596 &(6.2226,  &6.2177&0.0401&(6.2175, &\\
  && & & 6.2233) &&& 6.2180)&\\
 &CRR  &6.1912&&&6.2283&& & \\  
 &JR  &6.2281&&&6.2084&& & \\ 
\hline
\end{tabular}
\caption{Accuracy Comparison between MC-Tree and Binomial models in Pricing European Call Option. Bias: MC-Tree with the usage of Bias-Correction; Corr: MC-Tree with the usage of the distribution correction factor. }\label{mctreewithbinmodelscallopt}
\end{table}
\begin{table}
\centering
\begin{tabular}{|l|l|l|l|l|l|l|l|l|}
\hline
S & Method&
\multicolumn{3}{c|}{$N=50$}
&\multicolumn{3}{c|}{$N=100$}&
AS\\ 
 & &Mean&SD&CI & Mean&SD&CI  &  \\
\hline
100 &Corr &4.3720   &0.0324 &(4.3718,  &4.3720&0.0185&(4.3719, &4.3720\\
 & &  & &4.3722) &&&4.3721)&\\
  &Bias &4.3828  &0.0279 &(4.3827,  &4.3774 &0.0155&(4.3773, &\\
  & &  & & 4.3830 ) & &&4.3775  )&\\
 &CRR  &4.3657&&&4.3846&& & \\  
 &JR  &4.3600&&&4.3907&& & \\ 
\hline
90 &Corr &8.4048 &0.0503 &(8.4045,  &8.4048&0.0345&(8.4046, &8.4048\\
 & & & & 8.4051) &&& 8.4050)&\\
  &Bias &8.4153 &0.0596 &(8.4149,  &8.4101&0.0401&(8.4098, &\\
  && & & 8.4157) &&&8.4103 )&\\
 &CRR  &8.3835&&&8.4206&& & \\  
 &JR  &8.4204&&&8.4008&& & \\ 
\hline
\end{tabular}
\caption{Accuracy Comparison between MC-Tree and Binomial models in Pricing European Put Option. Bias: MC-Tree with the usage of Bias-Correction;  Corr: MC-Tree with the usage of the distribution correction factor. }\label{mctreewithbinmodelsputopt}
\end{table} 
It is evident from figure \ref{Fig5} that CRR and JR model is less stable and more volatile than the MC-Tree model as the tree depth increases.\\
Then, we can obtain the mean squared error (MSE) for different models from the range of tree depth from 1 to 100 in Table \ref{MSEtable}. MSE from the MC-Tree is the lowest among models.\\
\begin{table}[h]
\centering
\begin{tabular}{|l | l | l|l|}
\hline
 Models& MC-Tree&CRR&JR \\
\hline
MSE&0.00015354 &0.001074532 &0.001015309  \\
\hline
\end{tabular}
\caption{Mean-Squared Error for Various Models}\label{MSEtable}
\end{table} 
Figure \ref{PlotversusNandM} plots the European call prices with a range of M and the tree depth from 20 to 200. Clearly, the price is more stable with a rather large M.\\
\subsection{Pricing American Put Option}
We will present the numerical results of some experiments to the American Put option and compare them with the LSM and popular binomial tree models to gain some insight into the performance of the MC-Tree Method. The quadratic polynomial is used in the regression model \cite{Longstaff}.\\
All numerical experiments use the same parameters as mentioned in the previous section: pricing European option, except M=2000.
We consider various examples of American option valuation and compare our method with the LSM, CRR, and JR.
\subsubsection{Comparison to LSM Method}
We will compare the  standard deviation of the two methods to understand their accuracy. Table \ref{Table4} shows the mean and the standard deviation from simulation results with the initial stock price at 100, and the "true" price at 4.5415. The ”true” price of
an American put option is obtained by the convergent binomial method with the depth of tree at 50,000. \\
\begin{table}[h]
\centering
\begin{tabular}{|l | l | l|}
\hline
 & MC-Tree&LSM \\
\hline
Mean&4.5483& 4.5782 \\
\hline
Standard Deviation&0.0319&7.1828\\
\hline
\end{tabular}
\caption{Mean and  Standard Deviation of American Put Option.}\label{Table4}
\end{table}
As shown in the table \ref{Table4}, the  standard deviation of the LSM is much larger than the one of the MC-Tree method when their means are similar. The LSM needs a significant increase in the number of replications to improve accuracy, which leads to an increase in the computation time per simulation.\\
\begin{table}[h]
\centering
\begin{tabular}{|l | l | l|}
\hline
 & MC-Tree&LSM \\
\hline
Mean&  4.5483 & 4.5274  \\
\hline
Standard Deviation&0.0319&  0.8465 \\
\hline
Error&0.0068& 0.0141 \\
\hline
Computation Time (Seconds)& 10.4658 &10.5423  \\
\hline
M&2000&120000  \\
\hline
\end{tabular}
\caption{Mean and standard deviation of American put Option with similar running time.}\label{Table5}
\end{table}
It is evident from table \ref{Table5} that it is not sufficiently good for the LSM to obtain a small standard deviation as the MC-Tree method, even that the computation time of both are almost the same. It is concluded that the MC-Tree method provides us with more accuracy than the LSM at a similar computational cost.
\subsubsection{Comparison to Binomial Models}
\begin{figure}[ht!]
\begin{center}
\includegraphics[height=10.75cm,width=8.36cm]{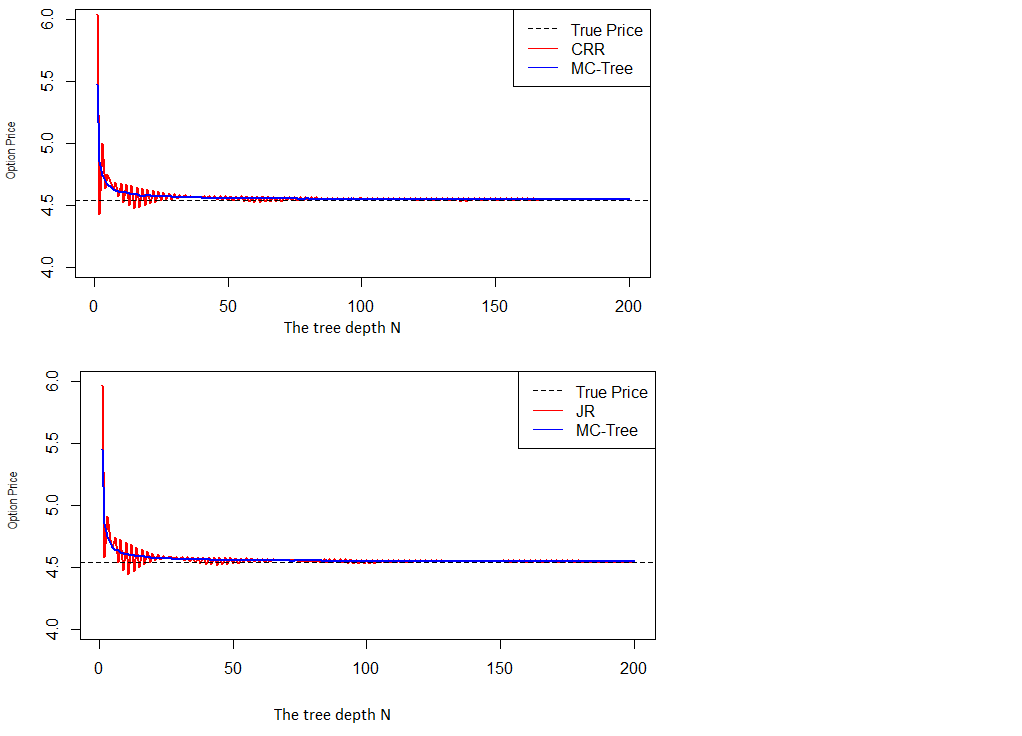}
\caption{Option prices v.s the tree depth N.}\label{Fig3}
\end{center}
\end{figure}

We will use the same model parameters, as mentioned in the previous section.
As shown in table \ref{Table6} and table \ref{Table7}, the option prices among models are insignificantly different, and the MC-Tree method produces the smallest error. It can be concluded that the MC-Tree method performs better than other methods: CRR, JR in terms of accuracy using the same tree-depth. Amin and Khanna \cite{Amin} proved that American option prices of the discrete model also converge to the corresponding value of the continuous-time model under fairly general conditions. It means that the "true" price can be obtained by increasing the tree depth to infinity. Therefore, we also compare results with the "true" prices.
\begin{table}[h]
\centering
\begin{tabular}{|l | l | l|l|l|l|}
\hline
Stock Price& MC-Tree&CRR&JR&"True" Price \\
\hline
95&6.4140& 6.3966& 6.4141&6.4058\\
\hline
97&5.6058&5.6148&5.6080 &5.5973  \\
\hline
100&4.5484 &4.5511&4.5583&4.5415   \\
\hline
102&3.9409& 3.9433&3.9240&3.9338  \\
\hline
104&3.4007&3.4034&3.4111&3.3960\\
\hline
\end{tabular}
\caption{Option Prices}\label{Table6}
\end{table}
\begin{table}[h]
\centering
\begin{tabular}{|l | l | l|l|l|}
\hline
 Stock Price& MC-Tree&CRR&JR \\
\hline
95&0.0081 & 0.0093&0.0083   \\
\hline
97 &0.0084 &0.0175&0.0107  \\
\hline
100&0.0080&0.0096&0.0168  \\
\hline
102&0.0070&0.0095& 0.0098\\
\hline
104&0.0047 &0.0073&0.0150\\
\hline
\end{tabular}
\caption{Accuracy Comparison among Models}\label{Table7}
\end{table}
Figure \ref{Fig3} shows two plots for prices of an American put option by the MC-Tree, CRR, and JR model when increasing the tree depth. The CRR and JR model both are more volatile than the MC-Tree model. The MC-Tree model is more stable and substantially less deviance from the overall downward, convergent toward the "true" price as N increases, compared with the CRR and JR model.\\
\subsection{Numerical Results of CVA Calculations}
The following parameters are used to estimate CVA of an American put option.\\
Initial stock price $S_0$=80, strike price K=100, expiration time T=1, risk-free rate r=0.03, volatility
$\sigma=0.2$, dividend rate=0, recovery rate R=0.4, intensity of default $\lambda=0.03$.\\
\begin{table}[h]
\centering
\begin{tabular}{|l | l |l | }
\hline
N&M&CVA \\
\hline
50&100&0.2447\\
\hline
75&150& 0.3013\\
\hline
100&250&0.3104 \\
\hline
250&700&0.3282 \\
\hline
250&10000&0.3392 \\
\hline
250&100000&0.3440 \\
\hline
2000&700&0.3414\\
\hline
4000&700&0.3414\\
\hline
\end{tabular}\\
\caption{CVA Values}\label{Table15}
\end{table}
CVA value is convergent to 0.34 when we increase the number of simulation M, and the tree depth N, respectively. 
\section{Concluding Remarks}
The method presents bounds on how close our results are to the "true" prices and shows the confidence interval containing the "true" one with a given (high) probability at 95\%.\\
Prices from MC-Tree method converge to the analytical solution or the "true" price. The completeness of the model allows to provide hedging strategies.\\
We implemented numerical experiments of MC-Tree on pricing options and CVA Calculations. We can conclude that MC-Tree method is an efficient method to price options on single asset. The model can be applied to practical development in financial industry due to its high accuracy.
\section{Further Research}
In the future, we intend to present the MC-Tree method, which combines the MC method with the recombining multinomial tree based on Pascal simplex \cite{Hanzon2018} for pricing multi-assets options. The research will be the natural generalization of the MC-Tree method in this article. \\
A future research direction is to generalize the model to a real market with stochastic parameters. Another promising research direction is to develop fast hardware implementations of the MC-Tree, which could be useful for the financial industry, especially the derivative pricing and risk management industry. For example, one can use an FPGA architecture to do the tree calculations, see for instance \cite{mahony2020parallel} where this is worked out for pricing multi-asset options. This could be combined with our MC approach to arrive at fast MC-Tree results. The difference with what is done in that paper would be in the precomputing phase, where we precompute the input parameters to the FPGA. 
The usage of the distribution correction factor brings very high accuracy in European option pricing using MC-Tree method. It will be interesting to investigate further how to use the distribution correction factor in pricing the American option.\\
\textbf{Acknowledgments}
We want to acknowledge Marta Ferrario and Luca Pettinari for their earlier contributions to the theory of the MC-Tree method in this paper.
\noindent

\nocite{*}
\bibliographystyle{plain}
\bibliography{Ref1.bib}
\end{document}